\documentclass{article}
\usepackage{arxiv}
\usepackage[utf8]{inputenc} 
\usepackage[T1]{fontenc}    
\usepackage{hyperref}       
\usepackage{url}            
\usepackage{booktabs}       
\usepackage{amsfonts}       
\usepackage{nicefrac}       
\usepackage{microtype}      
\usepackage{cleveref}       
\usepackage{graphicx}
\usepackage[numbers,sort&compress]{natbib}
\usepackage{doi}
\usepackage{authblk}        
\usepackage{float}          

\title{Facilitating Conditions as an Enabler, Not a Direct Motivator: A Robustness and Mediation Analysis of E-Learning Adoption}

\author[1]{Jaka Nugraha\thanks{\texttt{jakanugraha@unesa.ac.id}}}
\author[2]{Noyyn Sun\thanks{\texttt{noyny\_sun@ustc.edu}}}
\author[3]{Xinlin Zhao\thanks{\texttt{cillinzhao@gmail.com}}}
\author[1]{Vindi Kusuma Wardani\thanks{\texttt{vindikw09@gmail.com}}}
\author[4]{Inna Koblianska\thanks{\texttt{i.koblianska@biem.sumdu.edu.ua}}}
\author[5]{Jiunn-Woei Lian\thanks{\texttt{jwlian@nutc.edu.tw}}}

\affil[1]{Faculty of Economics and Business, Universitas Negeri Surabaya, Indonesia}
\affil[2]{Department of Economics, University of Amsterdam, Netherlands}
\affil[3]{Independent Researcher, China}
\affil[4]{Academic and Research Institute of Business, Economics, and Management, Sumy State University, Ukraine}
\affil[5]{Department of Information Management, National Taichung University of Science and Technology, Taiwan}

\renewcommand{\shorttitle}{}

\hypersetup{
pdftitle={Facilitating Conditions as an Enabler, Not a Direct Motivator},
pdfsubject={E-Learning Adoption},
pdfauthor={Nugraha et al.},
pdfkeywords={E-learning, UTAUT, learning engagement, digital platforms, e-learning efficiency},
}

\begin{document}
\maketitle

\begin{abstract}
Despite substantial institutional investment in e-learning infrastructure, student engagement often fails to meet expectations—a persistent paradox that challenges the established direct relationship between Facilitating Conditions (FC) and behavioral intention within the classic UTAUT framework. To resolve this theoretical puzzle, we reconceptualized the role of FC through an empirical study of 470 Indonesian university students. Our robust, multi-stage analytical approach first confirmed the significant influence of established drivers—Performance Expectancy ($\beta$=0.190), Effort Expectancy ($\beta$=0.198), Social Influence ($\beta$=0.151), and Perceived Enjoyment ($\beta$=0.472)—on Behavioral Intention (BI), which in turn strongly predicted Use Behavior ($\beta$=0.666). Crucially, however, the direct effect of FC on BI proved non-significant ($\beta$=-0.085). A subsequent mediation model revealed FC's true function as a foundational enabling construct that operates indirectly by powerfully enhancing both Performance Expectancy ($\beta$=0.556) and Effort Expectancy ($\beta$=0.419). Our findings demonstrate that the value of technological infrastructure lies not in its mere presence, but in its dynamic capacity to enable learning and optimize user experience. This research advances a refined "enabling pathway" theoretical framework, guiding administrators to shift the focus of technological investment from merely providing tools to strategically crafting learning experiences.

The research model incorporates key constructs, including the core UTAUT variables—Performance Expectancy (PE), Effort Expectancy (EE), Social Influence (SI), Facilitating Conditions (FC), Behavioral Intention (BI), and Actual Use Behavior (UB)—augmented with Personal Innovativeness (PI) and Perceived Enjoyment (PEJ). The structural relationships among these constructs are illustrated below:

\begin{figure}[H]
    \centering
    \caption{The Extended UTAUT Baseline Model. (This model confirms the significant effects of PE, EE, SI, and PEJ on BI, and BI on UB. However, the direct path from FC to BI (highlighted with a red dashed line) is not significant, presenting the core paradox of this study)}
    \includegraphics[width=0.8\textwidth, keepaspectratio]{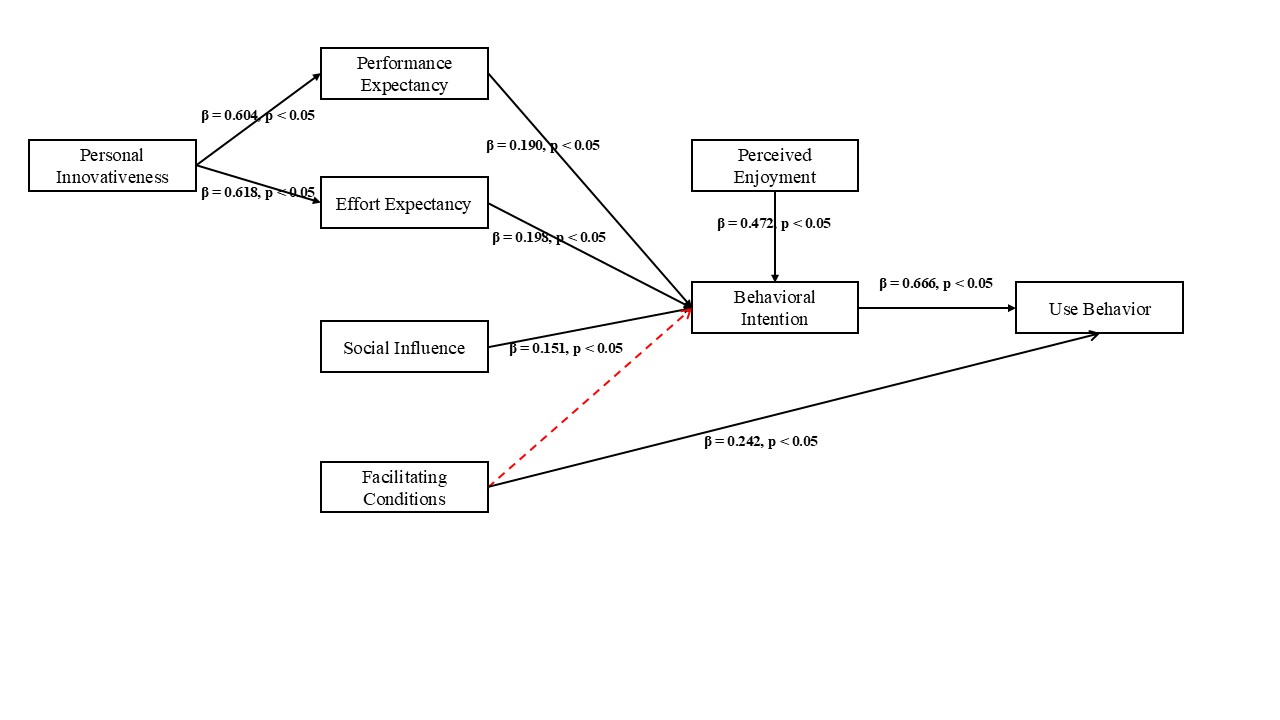}

    \label{fig:baseline_model}
\end{figure}

\begin{figure}[H]
    \centering
     \caption{The Mediation Model Elucidating the Indirect Mechanisms. (To resolve the paradox, this model proposes that FC influences BI indirectly by shaping learners' cognitive assessments. The model introduces two key mediating paths (highlighted in green bold lines) from FC to Performance Expectancy (PE) and Effort Expectancy (EE), which fully account for FC's effect on BI)}
    \includegraphics[width=0.8\textwidth, keepaspectratio]{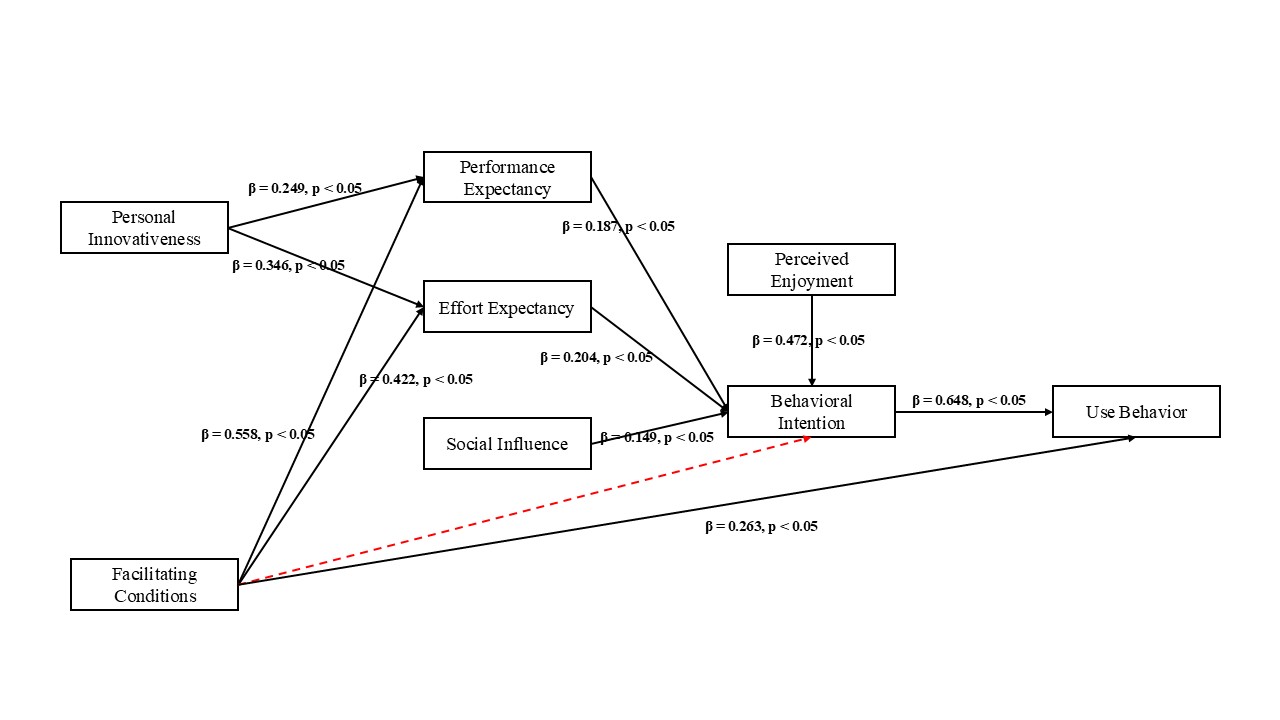}
   
    \label{fig:mediation_model}
\end{figure}

\end{abstract}


\section{INTRODUCTION}
The global higher education landscape has been profoundly reshaped by digital transformation, with E-Learning evolving from a supplementary tool to a core instructional mode (Ahmad et al., 2023). This shift, dramatically accelerated by the COVID-19 pandemic, has compelled institutions worldwide to invest heavily in digital learning infrastructure. A central premise behind these investments is the assumption, deeply embedded in foundational technology adoption models like the Unified Theory of Acceptance and Use of Technology (UTAUT), that providing robust Facilitating Conditions (FC)—such as reliable learning management systems, stable internet access, and technical support—directly encourages students to adopt and use these platforms (Venkatesh et al., 2003; Al-Fraihat et al., 2020).

However, a persistent and puzzling paradox often emerges in practice: despite substantial institutional investment, the adoption and sustained use of E-Learning platforms frequently fall short of expectations (Ahmad et al., 2023). This gap between resource allocation and actual usage is particularly salient in developing nations like Indonesia, where supportive policies like the "Kampus Merdeka" program coexist with challenges in digital literacy and infrastructure (Zulherman et al., 2021; BPS, 2023). This phenomenon challenges the conventional wisdom of a straightforward, direct link between FC and adoption intention (Behavioral Intention, BI).

A growing body of literature suggests that the role of FC may be more complex than traditionally modeled. While some studies affirm its direct impact, others in specific educational contexts find its direct effect on BI to be non-significant (e.g., Taamneh et al., 2022b), implying that its influence may be channeled through other perceptual factors. This inconsistency has prompted scholars to propose that FC may operate as an antecedent to core technology perceptions rather than a direct driver of intention. For instance, research has demonstrated that external factors often exert their influence indirectly by shaping perceived usefulness and ease of use (Al-Gahtani, 2016). This perspective is aligned with broader calls to re-examine and elaborate on the UTAUT's nomological network to better reflect complex realities (Dwivedi et al., 2019; Venkatesh et al., 2012).

Our investigation began with an empirical observation that directly contributes to this debate. We initially tested an extended UTAUT model on a sample of 470 Indonesian university students. The baseline model yielded a counterintuitive finding: while factors like Performance Expectancy (PE) and Perceived Enjoyment (PEJ) significantly predicted BI, the hypothesized direct path from FC to BI was not supported. A rigorous robustness check with 10,000 bootstrap iterations confirmed the stability of this anomalous result.

Faced with this robust paradox, we asked: If Facilitating Conditions do not directly motivate students, what is their true mechanism of action? Building upon prior theoretical suggestions (Al-Gahtani, 2016; Dwivedi et al., 2019), we proposed a novel mediation hypothesis: FC acts not as a direct motivator but as a foundational enabler. We theorized that its primary power lies in building the cognitive foundations for adoption by positively shaping students' core beliefs—specifically, by enhancing their Performance Expectancy (PE) and Effort Expectancy (EE).

To test this, we elaborated the baseline model into a new mediation model, introducing the pathways from FC to PE and EE. The results were striking The new model demonstrated a superior fit, and both new paths were highly significant (FC $\rightarrow$ PE: $\beta = 0.558$; FC $\rightarrow$ EE: $\beta = 0.422$ ), revealing FC's true influence as a critical antecedent to positive user perceptions. This study's pivotal contribution is the elaboration of this "enabling pathway"—a refined model that resolves the FC paradox by clarifying that infrastructure motivates not by its mere presence, but by making learning feel more effective and less effortful. For practitioners, it mandates a strategic pivot: to frame technological investments not as ends, but as means to enhance core user experiences.

Therefore, this study aims to clarify the true mechanism of Facilitating Conditions by contrasting the traditional direct-effect model with a newly proposed Infrastructure–Perception Bridge, testing whether FC influences Behavioral Intention indirectly through key cognitive beliefs.

\begin{figure}[H]
    \centering
    \caption{Theoretical Contrast: Traditional Model vs Proposed Infrastructure–Perception Bridge}
    \includegraphics[width=0.8\textwidth, keepaspectratio]{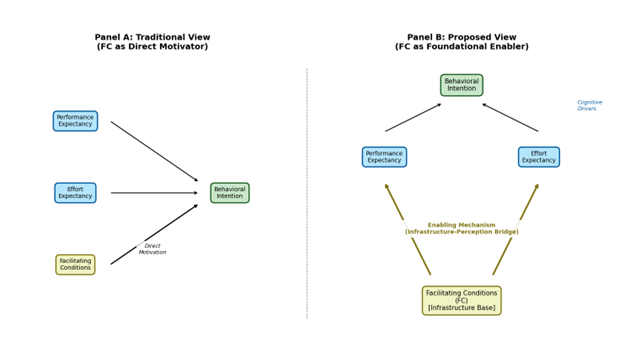}
    \label{fig:theoretical_contrast}
\end{figure}

\section{HYPOTHESIS DEVELOPMENT}

\subsection{Performance Expectancy (PE)}
PE is the level of confidence of a person in trusting the use of the system to successfully complete work and obtain good performance (Venkatesh et al., 2003). PE has a positive influence on user intentions to use technology in education (Lee, 2009); (Teo et al., 2019); (Chayomchai, 2020); (Taamneh et al., 2022b)). The global shift in education has accelerated the integration of digital technologies, leading to the rapid transformation of learning processes through e-learning in higher education. In this study, PE is the level of student confidence in using E-learning for e-learning to assist in the learning process. Based on this description, the first hypothesis is as follows.

H1: There is a significant positive effect between Performance Expectancy and student Behavioral Intention in using E-learning

\subsection{Effort Expectancy (EE)}
EE is defined as the level of ease associated with using the system (Venkatesh et al., 2003). Previous research stated that there was a positive effect of EE on intentions to use technology (Venkatesh et al., 2016). In the context of education, results have been proven regarding the positive influence of EE on users' intentions to use technology (Lee, 2009; Teo et al., 2019; Chayomchai, 2020; Taamneh et al., 2022). The implementation of e-learning in higher education has become an essential strategy to sustain and enhance the continuity and quality of the learning process in response to the growing demand for digital-based education. Based on this description, the second hypothesis is as follows.

H2: There is a significant positive effect between Effort Expectancy and student Behavioral Intention in using E-learning

\subsection{Social Influence (SI)}
SI is the extent to which a person perceives that the closest person believes that the person must use the new system (Venkatesh et al., 2003). Previous research has proven that there is an influence of SI on users' intention to use technology (Venkatesh et al., 2003). In the educational context, several researchers found SI’s direct positive effect on user intentions to use technology (Lee, 2009); McKenna et al., 2013; Oliveira et al., 2014; Teo et al., 2019; Chayomchai, 2020; Taamneh et al., 2022b). Taking into account the mandatory implementation of e-learning in higher education, further research is needed to determine the consistency of the research results. Therefore, SI in this study refers to the level of confidence students have in fellow students, lecturers and the university where they study to influence their intention to use E-learning. Based on this description, the third hypothesis is as follows.

H3: There is a significant positive influence between Social Influence and student Behavioral Intention in using E-learning

\subsection{Facilitating Conditions (FC)}
FC is defined as the degree to which a person believes the technical infrastructure provided by the organization is able to support the use of the system (Venkatesh et al., 2003). Previous research found a direct positive effect of FC on technology use behavior (McKenna et al., 2013; Oliveira et al., 2014; Venkatesh et al., 2003). In addition, several researchers found a significant positive effect between FC and intention to do something (Lee, 2009; Taamneh et al., 2022b). The rapid advancement of global digitalization has encouraged higher education institutions to adopt e-learning, although the effectiveness of its implementation in supporting learning outcomes still requires further evaluation. Based on this, the problem regarding the relationship between FC and the availability of facilities that support the implementation of e-learning will be consistent with previous research. In this study, FC refers to the extent to which students perceive that the university through the infrastructure provided influences them to use E-learning as an e-learning learning tool. Based on this description, the fourth and fifth hypotheses are as follows.

H4: There is a significant positive effect between Facilitating Condition and student Behavioral Intention in using E-learning

H5: There is a significant positive influence between students ' Facilitating Conditions and Actual Use Behavior in using E-learning

\subsection{Behavioral Intention (BI) and Actual Use Behavior}
BI has a strong influence on predicting a person's behavior in using technology (Venkatesh et al., 2003); (Al-Adwan et al., 2018; Lee, 2009; Taamneh et al., 2022b). Although several e-learning platforms have similarities in technology, users still need basic skills to operate these platforms using certain browsers, search engines, and operating systems. When compared to offline learning, e-learning users are required to take part in learning through certain websites or platforms by first registering to get an account and approval. Based on this description, the sixth hypothesis is as follows.

H6: There is a significant positive influence between students' Behavioral Intention and Actual Use Behavior in using E-learning

\subsection{Personal Innovativeness (PI)}
PI is defined as a psychological variable related to the behavior of users using technology so that they accept the technology (Sair \& Danish, 2018). Previous research found that PI has a positive and significant influence on technology’s performance expectancy and effort expectancy overall (Chayomchai, 2020; Sair \& Danish, 2018; Wu \& Lai, 2021) and in educational settings as well (Twum et al., 2022). In the context of e-learning, a person's innovation refers to the extent to which he is proficient in using the system. Innovative learners consider e-learning not a complex system to use so that it will influence them in using the system. PI in this study refers to the extent to which students have innovative abilities when using E-learning in the learning process. Based on this description, the seventh and eighth hypotheses are as follows.

H7: There is a significant positive effect between students' Personal Innovativeness and Performance Expectancy in using E-learning

H8: There is a significant positive effect between students' Personal Innovativeness and Effort Expectancy in using E-learning

\subsection{Perceived Enjoyment (PEJ)}
PEJ is defined as the extent to which a person's activity using a particular system is considered pleasurable in itself, regardless of the consequences resulting from using the system (Davis et al., 1992). Previous research found that there was a direct positive and significant relation between PEJ and behavioral intention (Chen et al., 2016; Lee, 2009; Pe-Than et al., 2015; Pipitwanichakarn \& Wongtada, 2021). The use of user-pleasing technology provides intrinsic motivation to use it. E-learning has various features that allow users to carry out various learning processes on one platform. Therefore PEJ will improve user attitudes to promote the use of e-learning. PEJ in this study refers to the extent to which students perceive E-learning as a fun platform in the learning process. Based on this description, the ninth hypothesis is as follows.

H9: There is a significant positive effect between Perceived Enjoyment and Student Behavioral Intention in using E-learning

\section{METHOD}
This research was conducted on students of the Faculty of Economics and Business, Universitas Negeri Surabaya. Students were asked to fill out an online questionnaire in May 2024. The population in this study were 2,765 active students of the Faculty of Economics and Business, State University of Surabaya, recorded in the Unesa Academic Information System (Siakadu) database. Sampling was done through purposive sampling technique, namely students who have used E-learning platform in learning process. Krecjie table (Bukhari, 2021) was used to determine the minimum sample size at a significance level of 0.05 (5\%), so that the minimum sample size for a population of 2,765 is 338, and this questionnaire was sent to 470 respondents.

\subsection{Data Collection and Analysis}
This study employed a cross-sectional survey design, with data collected through an online questionnaire platform (Google Forms). All measurement items were adapted from validated scales in existing literature and contextualized for the e-learning environment. Constructs were measured using a five-point Likert scale (1 = "Strongly Disagree" to 5 = "Strongly Agree"). The questionnaire underwent expert review and a pilot test (n=30) to ensure content validity, clarity, and contextual relevance. Cronbach's Alpha coefficients from the pilot test all exceeded 0.7, indicating good scale reliability.

The study population consisted of 2,765 active students enrolled in the Faculty of Economics and Business at Universitas Negeri Surabaya, Indonesia, during the 2023/2024 academic year. The minimum sample size, determined using the Krejcie \& Morgan (1970) table at a 95\% confidence level with a 5\% margin of error, was 338. A purposive sampling technique was employed, distributing the questionnaire link via official faculty channels to 470 students with prior experience using the official e-learning platform, ensuring respondents possessed the relevant background. Data collection was completed within May 2024.

A total of 430 questionnaires were returned. After screening for completeness and consistency, 12 invalid responses were excluded, yielding 418 valid questionnaires for final analysis, representing an effective response rate of 88.9\%. The demographic characteristics of the sample are presented in Table 6. Descriptive statistics indicate that the sample was 81.1\% female (n=339) and 18.9\% male (n=79); ages were concentrated between 19 and 21 years (84.8\%); and respondents came from eight different study programs, with Office Administration Education (26\%) and Economics (24.8\%) being the most represented, ensuring the sample's representativeness within the faculty.

\subsection{Data Analysis Strategy and Methods}
This study utilized Structural Equation Modeling - Generalized Structured Component Analysis as the primary data analysis technique. SEM-GSCA was selected over other SEM methods (e.g., CB-SEM, PLS-SEM) because it provides overall model fit indices akin to CB-SEM while enabling flexible hypothesis testing via bootstrapping, making it particularly suitable for exploratory model building and theory extension (Hwang, Cho, Jung, et al., 2021).

A sequential, three-stage analytical strategy was implemented to ensure robustness and probe underlying mechanisms deeply, as outlined in Figure 4.

\begin{figure}[H]
    \centering
    
    \caption{Three-stage data analysis process}
    \includegraphics[width=0.8\textwidth, keepaspectratio]{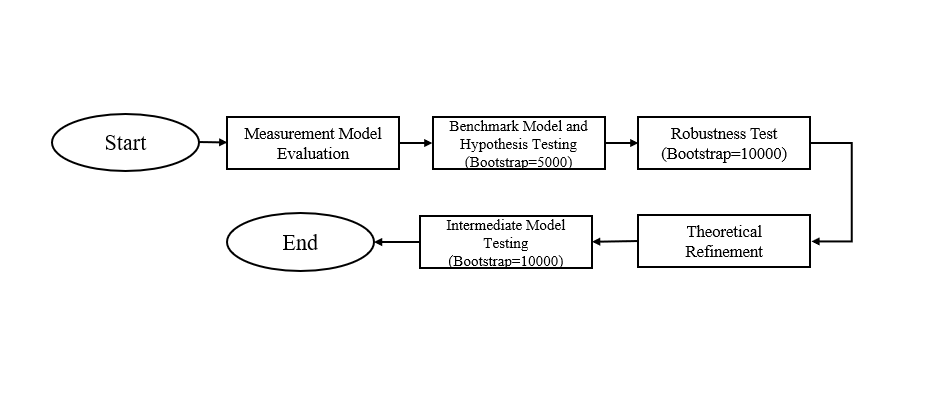}
    \label{fig:theoretical_contrast}
\end{figure}
Stage 1: Measurement Model Assessment

The reliability, convergent validity, and discriminant validity of the measurement model were rigorously assessed. As shown in Table 2, all constructs demonstrated composite reliability values between 0.770 and 0.961 and average variance extracted values between 0.694 and 0.891, exceeding the recommended thresholds of 0.70 and 0.50, respectively, indicating excellent internal consistency and convergent validity (Hair et al., 2014). Furthermore, as presented in Table 3, the square root of the AVE for each construct (diagonal values) was greater than its correlations with other constructs (off-diagonal values), providing sufficient evidence for discriminant validity (Fornell \& Larcker, 1981). The standardized factor loadings of all indicators on their respective constructs exceeded the recommended value of 0.70 (see Table 4), further confirming indicator validity. All variance inflation factor values were well below the threshold of 5.0 (see Table 5), indicating no severe multicollinearity concerns.

Stage 2: Baseline Model Testing and Robustness Assessment

After confirming the measurement model's quality, the baseline structural model (i.e., the initial extended UTAUT model) was estimated using 5,000 bootstrap samples to generate confidence intervals for path coefficients. This baseline model showed good fit: FIT = 0.677, AFIT =0.675, GFI = 0.992, SRMR = 0.049, with all indices meeting or exceeding recommended thresholds (Hwang, Cho, \& Choo, 2021). The results (see Table 8) indicated that most hypothesized paths were significant. However, the direct path from Facilitating Conditions to Behavioral Intention was non-significant ($\beta$ = -0.085, 95\% CI [-0.161, 0.002]), leading to the rejection of Hypothesis H4.

To verify the reliability of this key finding, a rigorous robustness check was conducted by increasing the bootstrap resamples to 10,000. The results showed that all parameter estimates, standard errors, and confidence intervals remained substantively unchanged (maximum change < 0.003). Crucially, the confidence interval for the FC $\rightarrow$ BI path remained inclusive of zero (95\% CI [-0.181, 0.011]) under 10,000 bootstraps. This strongly demonstrates that the rejection of H4 was not a statistical artifact but a highly stable finding.

Stage 3: Model Elaboration and Mediation Mechanism Test  

Guided by theoretical reasoning and the robust anomalous finding above, a competing mediation model was constructed and tested. This model introduced two new theoretical paths: FC $\rightarrow$ PE and FC $\rightarrow$ EE, to test the "enabling mechanism" of FC as an antecedent of core cognitive beliefs.

This mediation model demonstrated improved model fit (AFIT increased from 0.675 to 0.682). More importantly, as shown in Table 10, both newly added paths were highly significant and substantial: FC $\rightarrow$ PE ($\beta$ = 0.558, 95\% CI [0.460, 0.649]) and FC $\rightarrow$ EE ($\beta$ = 0.422 , 95\% CI [0.303, 0.535]). All analyses were performed using the GSCA Pro 1.1.1 software (Hwang, Cho, \& Choo, 2021).

\section{RESULTS}

\subsection{Measurement Model Assesment}

\begin{table}[H]
\centering
\caption{Indicators of Loading on Components}
\label{tab:loadings}
\begin{tabular}{lcccccccc}
\toprule
Indicator & PI & PE & EE & SI & FC & PEJ & BI & UB \\
\midrule
PE1 & 0.499 & \textbf{0.861} & 0.708 & 0.537 & 0.598 & 0.663 & 0.637 & 0.669 \\
PE2 & 0.518 & \textbf{0.882} & 0.689 & 0.570 & 0.571 & 0.669 & 0.681 & 0.689 \\
PE3 & 0.525 & \textbf{0.897} & 0.717 & 0.598 & 0.608 & 0.692 & 0.661 & 0.715 \\
PE4 & 0.556 & \textbf{0.903} & 0.730 & 0.552 & 0.652 & 0.711 & 0.668 & 0.712 \\
PE5 & 0.573 & \textbf{0.886} & 0.711 & 0.588 & 0.627 & 0.711 & 0.702 & 0.702 \\
BI1 & 0.587 & 0.722 & 0.716 & 0.614 & 0.568 & 0.751 & \textbf{0.933} & 0.744 \\
BI2 & 0.626 & 0.699 & 0.698 & 0.600 & 0.571 & 0.770 & \textbf{0.945} & 0.785 \\
BI3 & 0.619 & 0.725 & 0.723 & 0.602 & 0.566 & 0.769 & \textbf{0.954} & 0.769 \\
EE1 & 0.545 & 0.732 & \textbf{0.867} & 0.564 & 0.547 & 0.688 & 0.701 & 0.683 \\
EE2 & 0.539 & 0.699 & \textbf{0.892} & 0.557 & 0.508 & 0.643 & 0.643 & 0.651 \\
EE3 & 0.533 & 0.741 & \textbf{0.888} & 0.559 & 0.555 & 0.678 & 0.673 & 0.678 \\
EE4 & 0.540 & 0.598 & \textbf{0.832} & 0.486 & 0.461 & 0.590 & 0.602 & 0.609 \\
EE5 & 0.551 & 0.747 & \textbf{0.902} & 0.574 & 0.580 & 0.730 & 0.686 & 0.736 \\
FC1 & 0.536 & 0.689 & 0.666 & 0.565 & \textbf{0.797} & 0.688 & 0.620 & 0.661 \\
FC2 & 0.545 & 0.534 & 0.434 & 0.552 & \textbf{0.877} & 0.563 & 0.472 & 0.474 \\
FC3 & 0.537 & 0.558 & 0.464 & 0.627 & \textbf{0.893} & 0.562 & 0.464 & 0.524 \\
PEJ1 & 0.679 & 0.740 & 0.743 & 0.581 & 0.648 & \textbf{0.954} & 0.773 & 0.847 \\
PEJ2 & 0.669 & 0.744 & 0.729 & 0.617 & 0.673 & \textbf{0.947} & 0.767 & 0.843 \\
PEJ3 & 0.673 & 0.710 & 0.668 & 0.592 & 0.659 & \textbf{0.915} & 0.738 & 0.787 \\
PI1 & \textbf{0.825} & 0.583 & 0.606 & 0.497 & 0.604 & 0.697 & 0.604 & 0.643 \\
PI2 & \textbf{0.784} & 0.433 & 0.461 & 0.461 & 0.432 & 0.497 & 0.477 & 0.463 \\
PI3 & \textbf{0.871} & 0.495 & 0.481 & 0.478 & 0.508 & 0.585 & 0.534 & 0.567 \\
PI4 & \textbf{0.858} & 0.466 & 0.483 & 0.485 & 0.539 & 0.591 & 0.528 & 0.574 \\
PI5 & \textbf{0.886} & 0.538 & 0.539 & 0.457 & 0.543 & 0.619 & 0.563 & 0.602 \\
SI1 & 0.442 & 0.449 & 0.444 & \textbf{0.656} & 0.581 & 0.441 & 0.387 & 0.421 \\
SI2 & 0.484 & 0.546 & 0.550 & \textbf{0.921} & 0.556 & 0.549 & 0.581 & 0.548 \\
SI3 & 0.492 & 0.603 & 0.562 & \textbf{0.896} & 0.585 & 0.589 & 0.609 & 0.602 \\
UB1 & 0.605 & 0.749 & 0.736 & 0.592 & 0.610 & 0.830 & 0.741 & \textbf{0.929} \\
UB2 & 0.638 & 0.728 & 0.706 & 0.598 & 0.618 & 0.803 & 0.775 & \textbf{0.936} \\
UB3 & 0.663 & 0.722 & 0.697 & 0.580 & 0.565 & 0.825 & 0.748 & \textbf{0.925} \\
\bottomrule
\end{tabular}
\end{table}

Based on the table 1, it is known that the overall value of the indicator loading on component is greater than 0.7 or greater than the value of the other indicators. This model meets the requirements which requires the indicator value of loading on component is more than 0.7 (Hwang, Cho, Jung, et al., 2021) (Hair et al., 2014)(Meneau \& Moorthy, 2022)

\begin{table}[H]
\centering
\caption{Construct quality measures (Reliability of indicators)}
\label{tab:reliability}
\begin{tabular}{lcccccccc}
\toprule
 & PI & PE & EE & SI & FC & PEJ & BI & UB \\
\midrule
PVE & 0.715 & 0.785 & 0.769 & 0.694 & 0.734 & 0.882 & 0.891 & 0.866 \\
Alpha & 0.901 & 0.931 & 0.924 & 0.770 & 0.818 & 0.933 & 0.939 & 0.923 \\
rho & 0.926 & 0.948 & 0.943 & 0.869 & 0.892 & 0.957 & 0.961 & 0.951 \\
Dimensionality & 1 & 1 & 1 & 1 & 1 & 1 & 1 & 1 \\
\bottomrule
\end{tabular}
\end{table}

Construct Quality Measures (Reliability of Indicators) in table 2, the PVE values for the variables PI, PE, EE, SI, FC, PEJ, BI, and UB are more than 0.50. Then for the Alpha model on the variables PI, PE, EE, SI, FC, PEJ, BI, and, UB has a yield value of more than 0.70. Furthermore, in the Rho model for all variables, namely PI, PE, EE, SI, FC, PEJ, BI, and, UB has fulfilled the threshold value requirements, namely above 0.70. So all the variables in the construct quality measures (reliability of indicators) model have acceptable levels of convergent validity, internal consistency, and composite reliability. As in research conducted by (Hair et al., 2014) which suggested Alpha, and Rho values above 0.70 and PVE more than 0.50 (Ali et al., 2021). Dimensionality for the component is 1.00 (Meneau \& Moorthy, 2022).

The means, variance and R values of the components are shown in Table 3.

\begin{table}[H]
\centering
\caption{Component validity assessment}
\label{tab:validity}
\begin{tabular}{lcccccccc}
\toprule
Fornell-Larcker & & & & & & & & \\
criterion & PI & PE & EE & SI & FC & PEJ & BI & UB \\
values & & & & & & & & \\
\midrule
PI & 0.846 & & & & & & & \\
PE & 0.604 & 0.886 & & & & & & \\
EE & 0.618 & 0.802 & 0.877 & & & & & \\
SI & 0.565 & 0.642 & 0.625 & 0.833 & & & & \\
FC & 0.629 & 0.690 & 0.605 & 0.680 & 0.857 & & & \\
PEJ & 0.717 & 0.779 & 0.760 & 0.635 & 0.703 & 0.939 & & \\
BI & 0.647 & 0.757 & 0.754 & 0.641 & 0.602 & 0.809 & 0.944 & \\
UB & 0.682 & 0.787 & 0.766 & 0.634 & 0.643 & 0.879 & 0.812 & 0.930 \\
\bottomrule
\end{tabular}
\end{table}

Based on table 3 stated that All the off-diagonal values are less than that of average variance explained (AVE) by the first-order components indicating convergent validity (Fornell \& Larcker, 1981).

Further, the variance inflation factor (VIF) of components presented in table 4 is less than five indicating the absence of significant multi-collinearity issues (Hair et al., 2014).

\begin{table}[H]
\centering
\caption{Assessment of component correlation (VIF)}
\label{tab:vif}

\small

\setlength{\tabcolsep}{12pt} 

    \begin{tabular}{lccc} 
    \toprule
    Endogenous components & $\rightarrow$ & BI & UB \\
    
    Exogenous components & $\downarrow$ & & \\
    \midrule
    
    PI  & &       &       \\
    PE  & & 3.810 &       \\
    EE  & & 3.332 &       \\
    SI  & & 2.195 &       \\
    FC  & & 2.554 & 1.570 \\
    PEJ & & 3.348 &       \\
    BI  & &       & 1.570 \\
    \bottomrule
    \end{tabular}
\end{table}

\begin{table}[H]
\centering
\caption{R Square}
\label{tab:rsquare}
\begin{tabular}{cccccccc}
\toprule
PI & PE & EE & SI & FC & PEJ & BI & UB \\
\midrule
0 & 0.365 & 0.382 & 0 & 0 & 0 & 0.723 & 0.696 \\
\bottomrule
\end{tabular}
\end{table}

Based on the table 5, the BI variation is influenced by the independent variables in this research for 72.3\%, and the UB variation – for 69.6\%, meaning a good explanatory power of the model. For PE and EE variation, the model explains 36,5 \% and 38,2\%, respectively.

Overall, the measurement model meets the reliability, validity and collinearity tests, and proceed to assess the structural model.

\section{FINDINGS} 
The total number of respondents who filled out the survey distributed via the google form unesa.me/survey link accounts for 418 exceeding the minimum sample requirements.

The main data demonstrates that 81.1\% (339 respondents) are female while 18.9\% (79 respondents) are male. Furthermore, when viewed from the age range, it can be seen that 27 people aged 18 years (6.4\%), 129 people aged 19 years (30.8\%), 136 people aged 20 years (32.5\%), 90 people aged 21 years (21.5\%), 31 people aged 22 years (7.4\%), 5 people aged 23 years (1.4\%). While the review of the origin of the respondent's study program is as follows:

\begin{table}[H]
\centering
\caption{Respondents by study program}
\label{tab:respondents}
\begin{tabular}{clcc}
\toprule
No & Study program & Respondents & Respondents (\%) \\
\midrule
1 & Office Administration Education & 109 & 26 \\
2 & Economy & 104 & 24.8 \\
3 & Islamic economics & 66 & 15.7 \\
4 & Accounting education & 66 & 15.7 \\
5 & Accountancy & 32 & 7.6 \\
6 & Economic Education & 17 & 4 \\
7 & Business Education & 15 & 3.5 \\
8 & Management & 9 & 2.7 \\
\midrule
 & Total & 418 & 100 \\
\bottomrule
\end{tabular}
\end{table}

\subsection{Structural Model Assessment}

\begin{table}[H]
\centering
\caption{Structural model fit measures}
\label{tab:fitmeasures}
\begin{tabular}{cccccccc}
\toprule
FIT & AFIT & FITs & FITm & GFI & SRMR & OPE & OPEs \\
\midrule
0.677 & 0.675 & 0.270 & 0.785 & 0.992 & 0.049 & 0.324 & 0.733 \\
\bottomrule
\end{tabular}
\end{table}

Based on table 7, there is a FIT value of 0.677 indicating 67.7\% of the variance. The FITs value of 0.270 indicates that 27.0\% of the variance has been explained in the structural model. The FITm value is known to produce a value of 0.785, which means that the measurement model explains 78.5 \% of the variance. The estimated goodness of fit index (GFI) standardized root mean square residual (SRMR) value are 0.992 and 0.049, respectively. Meneau \& Moorthy (2022) stated that the estimate GFI value is greater than 0.93. Afterwards, the SRMR value has fulfilled the fit model requirements due to threshold value is 0.08 (Cho et al., 2020; Hwang, Cho, \& Choo, 2021).

\begin{table}[H]
\centering
\caption{Path Coefficient}
\label{tab:pathcoeff}

\small 

\setlength{\tabcolsep}{3.5pt} 

    \begin{tabular}{lccccccc}
    \toprule
    Path & Estimates & SE & 95\%CI(L) & 95\%CI(U) & Inferences & $f^2$ & \begin{tabular}[c]{@{}c@{}}Effect size\\ assessment\end{tabular} \\ 
    \midrule
    PE $\rightarrow$ BI & 0.190 & 0.059 & 0.091 & 0.314 & Accept H1 & 0.037 & small \\
    EE $\rightarrow$ BI & 0.198 & 0.049 & 0.100 & 0.304 & Accept H2 & 0.041 & small \\
    SI $\rightarrow$ BI & 0.151 & 0.044 & 0.071 & 0.247 & Accept H3 & 0.023 & small \\
    FC $\rightarrow$ BI & -0.085 & 0.044 & \textbf{-0.161} & \textbf{0.002} & \textbf{Reject H4} & 0.007 & small \\
    FC $\rightarrow$ UB & 0.242 & 0.042 & 0.171 & 0.327 & Accept H5 & 0.062 & small \\
    BI $\rightarrow$ UB & 0.666 & 0.040 & 0.583 & 0.732 & Accept H6 & 0.798 & large \\
    PI $\rightarrow$ PE & 0.604 & 0.036 & 0.504 & 0.659 & Accept H7 & 0.575 & large \\
    PI $\rightarrow$ EE & 0.618 & 0.029 & 0.570 & 0.667 & Accept H8 & 0.618 & large \\
    PEJ $\rightarrow$ BI & 0.472 & 0.053 & 0.355 & 0.572 & Accept H9 & 0.579 & large \\
    \bottomrule
    \end{tabular}
\end{table}

The GSCA Path Coefficient are presented in table 8. All paths are statistically significant based on at 95\% confidence intervals. The path of the structural equation model has a significant effect if the effect on the endogenous components is at 95\% confidence intervals and does not exceed zero. Performance Expectancy on Behavioral Intention has a path coefficient of 0.190 (CI L = 0.091 , CI U = 0.314), and the first hypothesis is accepted, which indicates that there is a positive effect of Performance Expectancy on Behavioral Intention. Effort Expectancy on Behavioral Intention has a path coefficient of 0.198 (CI L = 0.100, CI U = 0.304), and the second hypothesis is accepted. Social Influence on Behavioral Intention has a path coefficient of 0.151 (CI L = 0.071, CI U = 0.247), and the third hypothesis is accepted, which indicates that there is a positive effect of Social Influence on Behavioral Intention. Facilitating Condition on Behavioral Intention has a path coefficient of -0.085 (CI L = -0.161, CI U = 0.002), and the fourth hypothesis is rejected, due to includes zero which indicates that there is negative and no effect of Facilitating Condition on Behavioral Intention. Facilitating Conditions on Actual Use Behavior has a path coefficient of 0.242 (CI L = 0.171 , CI U = 0.327), and the fifth hypothesis is accepted, which indicates that there is a positive effect of Facilitating Conditions on Actual Use Behavior. Behavioral Intention to Actual Use Behavior has a path coefficient of 0.666 (CI L = 0.583, CI U = 0.732), and the sixth hypothesis is accepted, which indicates that there is a positive effect of Behavioral Intention on Actual Use Behavior. Personal Innovativeness on Performance Expectancy has a path coefficient of 0.604 (CI L = 0.504, CI U = 0.659), and the seventh hypothesis is accepted, which indicates that there is a positive influence of Personal Innovativeness on Performance Expectancy. Personal Innovativeness on Effort Expectancy has a path coefficient of 0.618 (CI L = 0.570, CI U = 0.667), and the eighth hypothesis is accepted, which indicates that there is a positive effect of Personal Innovativeness on Effort Expectancy. Perceived Enjoyment on Behavioral Intention has a path coefficient of 0.472 (CI L = 0.355, CI U = 0.572), and the ninth hypothesis is accepted, which indicates that there is a positive effect of Perceived Enjoyment on Behavioral Intention. This table shows the f2 effect size of each predictor component. As a rule of thumb, the f2 values of 0.02, 0.15 and 0.35 may be considered small, medium, and large effect sizes, respectively (Cohen, 1988).

\subsection{Robustness Testing Analysis}
To ensure the reliability of our initial findings—particularly the counterintuitive lack of a direct effect between Facilitating Conditions (FC) and Behavioral Intention (BI)—we conducted a robustness check by increasing the bootstrap resampling from 5,000 to 10,000 iterations. This analysis served to determine whether the key findings were susceptible to random sampling variation. The results demonstrated remarkable stability across all key parameters. Crucially, the path coefficient for H4 (FC $\rightarrow$ BI) remained non-significant, with its confidence interval consistently containing zero. Conversely, all paths that were statistically significant in the baseline model retained their significance.

Table 9
\begin{table}[H]
\centering
\caption{Robustness Check: Comparison of Path Coefficients (Bootstrap 5,000 vs 10,000)}
\label{tab:robustness}

\small

\setlength{\tabcolsep}{8pt}

    \begin{tabular}{lccccc}
    \toprule
    Path & 
    \begin{tabular}[c]{@{}c@{}}Estimates\\ (5,000)\end{tabular} & 
    \begin{tabular}[c]{@{}c@{}}Estimates\\ (10,000)\end{tabular} & 
    SE(5,000) & 
    SE(10,000) & 
    Inference \\
    \midrule
    
    PE $\rightarrow$ BI  & 0.19   & 0.191  & 0.059 & 0.059 & Significant \\
    EE $\rightarrow$ BI  & 0.198  & 0.199  & 0.049 & 0.052 & Significant \\
    SI $\rightarrow$ BI  & 0.151  & 0.152  & 0.044 & 0.044 & Significant \\
    FC $\rightarrow$ BI  & -0.085 & -0.085 & 0.044 & 0.049 & Not Significant \\
    FC $\rightarrow$ UB  & 0.242  & 0.242  & 0.042 & 0.044 & Significant \\
    BI $\rightarrow$ UB  & 0.666  & 0.666  & 0.04  & 0.04  & Significant \\
    PI $\rightarrow$ PE  & 0.604  & 0.605  & 0.036 & 0.035 & Significant \\
    PI $\rightarrow$ EE  & 0.618  & 0.618  & 0.029 & 0.032 & Significant \\
    PEJ $\rightarrow$ BI & 0.472  & 0.473  & 0.053 & 0.054 & Significant \\
    \bottomrule
    \end{tabular}
\end{table}

The changes in parameter estimates, standard errors, and confidence interval boundaries were negligible (all variations < |0.003|), well within the range of sampling error. This provides strong evidence that the core conclusions of our baseline model are robust and not an artifact of the bootstrap specification.

\subsection{Exploration and Analysis of Intermediary Models}
The statistically non-significant path between Facilitating Conditions (FC) and Behavioral Intention (BI) in the baseline model presented a theoretically meaningful anomaly. To investigate the underlying mechanisms, we formulated a mediation model positing that FC influences BI indirectly through cognitive pathways. Specifically, we hypothesized that adequate institutional support enhances users' perceptions of system usefulness (Performance Expectancy, PE) and reduces perceived complexity (Effort Expectancy, EE), which subsequently shape behavioral intentions.

Initially, the absence of a significant direct effect between Facilitating Conditions (FC) and behavioral intention (BI) prompted us to improve the model to elucidate its underlying causal mechanisms:

Theoretical Basis: We hypothesize that the primary function of Facilitating Conditions (FC) is to establish preconditions conducive to product adoption by positively influencing users' key cognitive appraisers—namely, performance expectation (PE) and effort expectation (EE).

Methodological Implementation: To test this mediation model, we introduce theoretical paths from facilitating conditions (FC) to performance expectation and from facilitating conditions (FC) to effort expectation (EE). This setting allows us to explicitly assess the indirect effect of facilitating conditions (FC) on behavioral intention through two hypothesized mediating variables.

Our comparative analysis of the fit indices of the baseline and mediation models, along with a compelling narrative, demonstrates the shift from empirical anomalies to theoretical explanations. By systematically integrating model fit indices, critical path coefficients, and structural relationships, we demonstrate that the mediation model is theoretically superior and provides a more nuanced interpretation of the observed data.

\begin{table}[H]
\centering
\caption{Mediation Model Fit Indices Comparison}
\label{tab:mediationfit}

\small

\renewcommand{\arraystretch}{1.5}

\setlength{\tabcolsep}{4pt}

    \begin{tabular}{lcc p{4cm} p{5.8cm}}
    \toprule
    \textbf{Fit Index} & 
    \textbf{\begin{tabular}[c]{@{}c@{}}Baseline\\ Model\end{tabular}} & 
    \textbf{\begin{tabular}[c]{@{}c@{}}Mediation\\ Model\end{tabular}} & 
    \textbf{Threshold} & 
    \textbf{Interpretation} \\
    \midrule

    FIT & 
    0.677 & 
    0.684 & 
    Ranges from 0 to 1 \newline (Values closer to 1 indicate better variance explanation) & 
    The mediation model demonstrates marginally improved explanatory power, accounting for a greater proportion of the total variance. \\

    AFIT & 
    0.675 & 
    0.682 & 
    Higher values are preferred (Primary criterion for model comparison) & 
    Critical Metric. The elevated AFIT in the mediation model indicates that the inclusion of additional paths justifies the increased model complexity, signifying overall model superiority. \\

    GFI & 
    0.992 & 
    0.994 & 
    $\ge 0.90$ (Acceptable) \newline $\ge 0.93$ (Good) & 
    Both models substantially exceed the recommended threshold, indicating excellent goodness-of-fit; the mediation model exhibits a slight advantage. \\

    SRMR & 
    0.049 & 
    0.046 & 
    $\le 0.08$ (Good fit) & 
    Both values remain well below the cutoff, implying minimal standardized residuals; the mediation model demonstrates superior predictive accuracy. \\

    \bottomrule
    \end{tabular}
\end{table}

The analysis of overall model fit provides the foundational evidence for our theoretical elaboration. As demonstrated in the comparison of fit indices, the mediation model shows a superior alignment with the empirical data. Specifically, the decrease in SRMR and the increase in the AFIT value are particularly critical. The improvement in AFIT is a strong indicator that the enhancement in model explanatory power outweighs the penalty for increased complexity introduced by the additional paths. This confirms that the mediation model is not merely an alternative specification, but a theoretically more adequate representation of the underlying phenomena.

\begin{table}[H]
\centering
\caption{New intermediary path verification: reveal the real mechanism of FC}
\label{tab:newpaths}
\begin{tabular}{lccc}
\toprule
New path & Estimates & 95\%CI (Lower, Upper) & Conclusion \\
\midrule
FC $\rightarrow$ PE & 0.558 & [0.460, 0.649] & Significant \\
FC $\rightarrow$ EE & 0.422 & [0.303, 0.535] & Significant \\
\bottomrule
\end{tabular}
\end{table}

The validation of the newly added paths—FC $\rightarrow$ PE and FC $\rightarrow$ EE—represents the core of our theoretical contribution. The strong, significant coefficients for these paths ($\beta$=0.558 and $\beta$=0.422, respectively) force a fundamental reconceptualization of the Facilitating Conditions construct. FC’s role is not that of a direct motivator but that of a critical enabler. It functions by building what we term the "Infrastructure-Perception Bridge," shaping students' core cognitive assessments. Robust technological infrastructure and support (high FC) directly lead to the perception that the e-learning system is more useful (higher PE) and easier to use (lower EE). This finding resolves the initial paradox by shifting the question from if FC matters to how it matters.

\begin{table}[H]
\centering
\caption{Mediation Model Path Coefficients}
\label{tab:mediationcoeff}
\begin{tabular}{lccccl}
\toprule
Path & Estimates & SE & 95\%CI Low & 95\%CI Up & Inference \\
\midrule
PI $\rightarrow$ PE & 0.249 & 0.049 & 0.152 & 0.347 & Significant \\
FC $\rightarrow$ PE & 0.558 & 0.048 & 0.460 & 0.649 & Significant \\
PI $\rightarrow$ EE & 0.346 & 0.055 & 0.240 & 0.456 & Significant \\
FC $\rightarrow$ EE & 0.422 & 0.059 & 0.303 & 0.535 & Significant \\
PE $\rightarrow$ BI & 0.187 & 0.059 & 0.070 & 0.302 & Significant \\
EE $\rightarrow$ BI & 0.204 & 0.053 & 0.099 & 0.306 & Significant \\
SI $\rightarrow$ BI & 0.149 & 0.043 & 0.065 & 0.237 & Significant \\
FC $\rightarrow$ BI & -0.082 & 0.050 & -0.179 & 0.019 & Not Significant \\
FEJ $\rightarrow$ BI & 0.472 & 0.054 & 0.366 & 0.578 & Significant \\
FC $\rightarrow$ UB & 0.263 & 0.046 & 0.174 & 0.356 & Significant \\
BI $\rightarrow$ UB & 0.648 & 0.042 & 0.562 & 0.727 & Significant \\
\bottomrule
\end{tabular}
\end{table}

The changes in the previously established paths further corroborate the validity of the mediating model. The most telling alteration is the marked decrease in the influence of Personal Innovativeness (PI) on both PE and EE. This indicates that in the baseline model, PI was likely accounting for variance that is more accurately explained by the shared influence of FC. The revised model presents a more complete story: students' positive cognitive assessments are a function of both their innate innovativeness (PI) and the enabling environment (FC). Furthermore, the strengthening of the path from BI to UB suggests that by more accurately specifying the antecedents of behavioral intention, the intention itself becomes a sharper, more reliable predictor of actual behavior.

\section{DISCUSSIONS} 
\subsection{Performance Expectancy and Behavioral Intention}
The results of this study support previous findings which state that the use of technology through a variety of feature designs makes users able to access it and encourages them to use the system (Lee, 2009). In addition, technology has made an organization always give priority when developing technology by considering the characteristics of users. Thus, the developed system can be utilized optimally by users in supporting their activities (Chayomchai, 2020; A. Taamneh et al., 2022a; Teo et al., 2019).

\subsection{Effort Expectancy and Behavioral Intention}
This research was supporting previous research which indicates that effort expectancy affects student behavioral intention when using e-learning (Lee, 2009; Teo et al., 2019; Chayomchai, 2020; Taamneh et al., 2022). The results of the study show that E-learning has a high level of ease when students do learning. There are several features that make it easy for students to carry out activities in collecting assignments and responding to material provided directly through the forum feature. Integration of learning systems with smartphones is also the reason for using this platform which has a high level of convenience.

\subsection{Social Influence and Behavioral Intention}
The results in this study support the findings which state that the closest people have given confidence to use technology and use it as a form of necessity in carrying out their activities (Oliveira et al., 2014). The use of e-learning has encouraged the implementation of learning to be carried out online which indirectly affects social interaction in education. The influence of student involvement when using technology and instructions from lecturers has made students use learning activities through E-learning.

\subsection{Facilitating Condition, Behavioral Intention and Actual Use Behavior}
Services provided through technical infrastructure that cause the use of technology in activities have made users feel comfortable and provide convenience when interacting with this technology (Lee, 2009; Taamneh et al., 2022). The research results have shown consistency with previous findings which state that the availability of facilities that support the implementation of e-learning has helped users use technology (Venkatesh et al., 2003; McKenna et al., 2013; Oliveira et al., 2014). Use Behavior in the use of e-learning provides features that help system providers make improvements to technical errors to ensure user ease in using the system by providing e -service features in offering assistance.

\subsection{Behavioral Intention and Actual Use Behavior of students}
Technology in e-learning has required users to be familiar with various features related to activities such as offline learning. Thus, users are required to be able to operate e-learning devices using web browsers and the like. The results of this study consistently support previous findings which state that users are required to follow certain mechanisms and stages in utilizing e-learning, including how to use various features provided by system providers so that e-learning can be more optimal (Lee, 2009; Al- Adwan et al., 2018; Taamneh, 2022).

\subsection{Personal Innovativeness Effort Expectancy and Student Performance Expectancy}
Innovative learners will be able to take advantage of the features provided in e-learning systems which have an impact on increasing individual performance in the learning process. User-friendly features have helped students become more proficient and current when using systems that have been complexly created by developers. The research results have supported previous findings which stated that individual user skills and innovation had an impact on optimal performance and made it easier for system users to behave with technology ( Sair and Danish, 2018: Chayomchai, 2020; Wu and Lai, 2021)

\subsection{Perceived Enjoyment and Behavioral Intention of students}
Technology often has rich entertainment functions, and users can get great pleasure when playing it (Lee, 2009). Users feel that a technology produces more pleasure during use, so users will feel that the technology is easy to use which causes users to become more familiar and gain more understanding including technological capabilities after a period of use (Pipitwanichakarn \& Wongtada, 2019). The results of this study are consistent with previous findings which state that through a variety of features provided by technology developers, users are motivated to continue using and encourage themselves to be actively and innovatively involved in fun learning (Lee, 2009; Chen et al., 2015; Than et al 2015; Pipitwanichakarn et al., 2019).

This study aimed to investigate the determinants of e-learning adoption in Indonesian higher education. While the initial baseline model confirmed the significance of most UTAUT constructs, it revealed a critical anomaly: Facilitating Conditions (FC) had no direct significant effect on Behavioral Intention (BI). By restructuring the model into a full mediation framework, we successfully resolved this paradox. The discussion below focuses on this theoretical refinement and the re-evaluation of FC's role.

\subsection{The Paradox of Facilitating Conditions: From Direct Driver to Foundational Enabler}
The rejection of Hypothesis H4 (FC $\rightarrow$ BI) in our baseline model challenges the original UTAUT proposition that organizational and technical infrastructure directly compels usage intention. In the context of Indonesian higher education, our findings suggest that the mere availability of hardware, internet, or helpdesks does not directly spark the intent to learn. Instead, the strong validation of the mediation paths (FC $\rightarrow$ PE and FC $\rightarrow$ EE) reveals that FC operates as a distal antecedent rather than a proximal determinant.

This supports the "Infrastructure-Perception Bridge" conceptualization. Adequate facilitating conditions function by removing barriers and lowering the cost of entry. When infrastructure is robust (High FC), students perceive the system as requiring less effort (High EE) and, consequently, more capable of delivering academic results (High PE). Conversely, without this foundation, students' cognitive assessment of the system's utility and ease of use collapses, inhibiting adoption. Thus, facilitating conditions (FC) acts as an enabler that activates the cognitive pathways (PE and EE), rather than a motivator that directly drives intention.

\subsection{Model Comparison and Theoretical Superiority}
The comparative analysis of model fit indices provides robust empirical support for this theoretical shift. Although the baseline model demonstrated acceptable fit, the mediation model outperformed it across key metrics. Most notably, the Adjusted Fit Index (AFIT) increased from 0.675 to 0.682. Since AFIT penalizes model complexity, this increase signifies that the explanatory power gained by adding the mediation paths far outweighs the cost of the added complexity. Additionally, the reduction in SRMR (from 0.049 to 0.046) indicates a more precise prediction of the observed data covariance. These statistical improvements confirm that the full mediation model provides a more accurate representation of the reality of e-learning adoption, emphasizing the indirect but indispensable role of institutional support.

\subsection{The Roles of Enjoyment and Innovativeness}
Consistent with recent extensions of technology acceptance research, Perceived Enjoyment (PEJ) emerged as a dominant predictor of intention, highlighting the importance of intrinsic motivation in voluntary learning environments. Furthermore, the refined model clarified the role of Personal Innovativeness (PI). In the mediation model, the influence of PI on PE and EE remained significant, but it was adjusted compared to the baseline, suggesting that a student's cognitive beliefs are shaped by a dual process: their internal trait (PI) and the external environment (FC).

\section{CONCLUSION}
\subsection{Summary of Findings and Theoretical Contribution}
This study set out to resolve a persistent paradox in educational technology adoption: the frequent disconnect between substantial investments in Facilitating Conditions (FC) and student engagement levels. Through a robust, multi-stage analysis of data from Indonesian university students, we demonstrated that the conventional view of FC as a direct motivator is incomplete. Instead, FC functions as a foundational enabler. Its influence is fully mediated by the learners’ core cognitive assessments—namely, Performance Expectancy (PE) and Effort Expectancy (EE). This discovery led us to propose the “Infrastructure-Perception Bridge” as a key mechanism and to advance a refined “enabling pathway” framework within the UTAUT model. Our primary theoretical contribution lies in this reconceptualization, which clarifies the nomological network by specifying that technological infrastructure drives adoption not by its mere presence, but by its capacity to make learning feel demonstrably more effective and less effortful.

\subsection{Practical Implications}
The“enabling pathway” framework mandates a strategic pivot for educators and policymakers, from a tool-centric to an experience-centric paradigm:

1. Shift from Hardware to Experience Design: Institutional investment must transcend infrastructure provisioning to prioritize intuitive Learning Experience (LX) design. Simplifying platform workflows and interfaces transforms technological capacity into tangible user value.

2. Engineer for Effortlessness: E-learning systems should be designed to minimize cognitive load through features like contextual help and intelligent navigation, directly enhancing Effort Expectancy (EE), a critical driver of adoption.

3. Make Value Visible: The performance benefits of robust infrastructure should be explicitly communicated to learners via analytics and feedback, thereby strengthening Performance Expectancy (PE) and linking support directly to perceived academic success.

\subsection{Limitations and Future Research Directions}
While this study provides compelling evidence for the“enabling pathway,” its findings should be considered in light of certain limitations, which also chart a course for future inquiry.

Contextual Generalizability: Our sample, drawn from a specific faculty in one Indonesian university, suggests caution in generalizing the results. Future research should validate this model across diverse cultural, national, and disciplinary contexts to establish its boundary conditions and enrich our understanding of contextual dependencies.

Model Comprehensiveness: Although PE and EE are powerful mediators, the adoption process is multifaceted. Future theoretical work could integrate additional psychological constructs—such as habit, flow, or perceived risk—to develop a more holistic model of technology adoption.

Causality and Dynamics: The cross-sectional design, despite robust analytical checks, limits causal claims. Longitudinal studies and field experiments are needed to establish causality definitively and to uncover how the “enabling pathway” evolves over time as users gain experience and as technological systems develop.

In summary, this study reconceptualizes the role of Facilitating Conditions, revealing it as a foundational enabler rather than a direct motivator. The proposed “Infrastructure-Perception Bridge” and the “enabling pathway” framework advance UTAUT theory by clarifying how infrastructure translates into adoption. Practically, it guides a crucial shift from deploying technology to designing experiences, ensuring digital learning environments are not only available but are intrinsically valued by learners for their utility and ease of use.

\end{document}